\documentclass[%
 reprint,
 amsmath,amssymb,
 aps,
]{revtex4-2}

\usepackage{graphicx}
\usepackage{dcolumn}
\usepackage{bm}
\usepackage{lipsum}
\usepackage{amsfonts}
\usepackage{amssymb}
\usepackage[latin1]{inputenc}
\usepackage{textcomp}
\usepackage{color}
\newcommand{\bra}[1]{\langle #1 \vert}
\newcommand{\ket}[1]{\vert #1 \rangle}
\newcommand{\norma}[1]{\vert #1 \vert^{2}}
\definecolor{darkspringgreen}{rgb}{0.09, 0.45, 0.27}

\begin{document}



\title{Noise correlations behind superdiffusive quantum walks}

\author{Gra\c{c}a R. M. de  Almeida,  N. Amaral,  A. R. C. Buarque and W. S. Dias}
\address{Instituto de F\' isica, Universidade Federal de Alagoas, 57072-900 Macei\' o, Alagoas, Brazil.}

\begin{abstract}
We study how discrete-time quantum walks behave under short-range correlated noise. By considering noise as a source of inhomogeneity of quantum gates, we introduce a primitive relaxation in the assumption of uncorrelated stochastic noise: binary pair correlations manifesting in the random distribution. Using different quantum gates, we examined the transport properties for both spatial and temporal noise regimes. For spatial inhomogeneities, we unveil noise correlations driving quantum walks from the well-known exponentially localized regime to superdiffusive spreading. This scenario displays an intriguing performance in which the superdiffusive exponent is almost invariant to the degree of inhomogeneity. The time-asymptotic regime and the finite-size scaling also unveil an emergent superdiffusive behavior for quantum walks undergoing temporal noise correlation, replacing the diffusive regime exhibited when noise is random and uncorrelated. However, some quantum gates are insensitive to correlations, contrasting with the spatial noise scenario. Numerical and analytical results provide valuable insights into the underlying mechanism of superdiffusive quantum walks, including those arising from deterministic aperiodic inhomogeneities.
\end{abstract}


\maketitle

\section{introduction}
\label{introduction}

Dynamical aspects of particles in discrete systems are among the fundamental issues in physics, the application of which extends to a wide variety of systems such as behavioral macroeconomics~\cite{NELSON1982139}, image segmentation~\cite{1704833}, animal dynamics~\cite{Viswanathan1996,Codling2008}, computer science\cite{10.1007/11569596_31}, evolutionary ecology ~\cite{Dieckmann1996,METZ1992198} and thermal conductivity of nanofluids~\cite{KEBLINSKI2002855,doi:10.1063/1.1756684}. The emergence of quantum walks further extends its importance, whether through a universal model for quantum computing~\cite{PhysRevA.48.1687,PhysRevLett.102.180501,PhysRevA.58.915,PhysRevA.81.042330} and the development of new quantum algorithms~\cite{doi:10.1142/S0219749903000383} but also for providing a versatile and highly controllable platform to describe quantum systems~\cite{PhysRevA.72.062317,PhysRevB.76.155124,Peruzzo1500,Preiss1229,PhysRevLett.106.180403,crespi2013,VenegasAndraca2012}.

Designing and controlling such quantum processes for long-time dynamics is essential, with noise among the principal obstacles~\cite{RevModPhys.89.035002,PhysRevA.94.052325,Campbell2017}. Quantum error-correcting methods~\cite{PhysRevA.52.R2493,PhysRevLett.77.793} and fault-tolerant protocols~\cite{10.1145/258533.258579,Knill342} have pointed to the need for a better understanding of the noise nature of the system. Thus, ingredients that symbolize interaction between system and environment have been studied. In general, noise drives the discrete-time quantum walk at a slower spreading rate in the long-time limit. White noise fluctuations on the time evolution operator usually lead to a diffusive wave-function spreading~\cite{PhysRevA.68.062315,PhysRevA.74.022310,PhysRevLett.106.180403,Joye2011,DiMolfetta2016}, while an arbitrary spatial inhomogeneity is responsible for a localized behavior~\cite{PhysRevLett.106.180403,crespi2013,PhysRevA.89.022309,Konno2010,PhysRevA.85.012329}. Studies also contemplate the coexistence of both scenarios, wherein the diffusive behavior over a long time limit has been documented~\cite{Ahlbrecht2012,PhysRevA.89.042307}. The consequences of an instantaneous stochastic noise over the quantum walk stability have been recently reported, where a maximally coherent initial state achieves breathing dynamics or even a standing self-focusing in a long-time regime~\cite{PhysRevA.103.042213}.

Noise correlations have attracted significant attention by quantum characterization, verification, and validation techniques~\cite{Bylander2011,PhysRevA.93.022303,Mavadia2018,Cai2020}. This aspect has also been considered by quantum walks, with reports of significant changes in the walk profile. For example, we observe a quantum walker exhibiting a superdiffusive spreading in one-dimensional systems where quantum gates follow aperiodic time-dependent sequences, such as Fibonacci~\cite{PhysRevLett.93.190503,PhysRevE.96.012111} and Thue-Morse~\cite{PhysRevE.96.012111}. Conversely, systems where quantum gates are temporally controlled based on the Rudin-Shapiro distribution exhibit a nearly diffusive behavior~\cite{PhysRevE.96.012111}. The observed behavior in the system utilizing Fibonacci temporal sequencing is connected with the power-law decay of the time-correlation function of the trace map~\cite{ROMANELLI20093985}. 

The nonstochastic scenario has also been explored in the spatial framework for different quantum gates distributed along the lattice sites and for systems with position-dependent phase defects. Aperiodic Fibonacci and Thue-Morse sequencing show a superdiffusive spreading, either embedded into the quantum gate distribution~\cite{PhysRevE.96.012111} or the step operator, where the allowed jumps symbolize connections between non-neighboring quantum gates~\cite{PhysRevE.102.012104}. We observe spatial localization of quantum walker for systems in which quantum gates are distributed analogous to the Aubry-Andr\'e model~\cite{PhysRevE.82.031122} and the Rudin-Shapiro ordering~\cite{PhysRevE.96.012111}. Transitions between localized and delocalized spreading were reported for systems with quantum gates following spatial aperiodicity~\cite{PhysRevE.100.032106} and systems with long-range correlations encoded as static phase disorder in the conditional shift operator~\cite{PhysRevE.99.022117}. Spatial inhomogeneity has been explored considering a hierarchical arrangement of barriers, suggesting a regime where quantum walks never appear to be localized~\cite{PhysRevResearch.2.023411}. Long-range correlations in systems with inhomogeneous space and time have shown a wide range of dynamic regimes, from localized to ballistic spreading~\cite{PhysRevE.107.064139}.


The reports illustrate how studies into inhomogeneous quantum walks focus on uncorrelated heterogeneities, which could arise from a stochastic noise, and heterogeneities following deterministic sequences or exhibiting long-range correlations. Less extreme and more realistic scenarios need further understanding. How do discrete-time quantum walks behave under short-range correlated noises? In this paper, we examine the impact of a primitive relaxation in uncorrelated stochastic noise assumption: the emergence of binary pair-correlated in the random distribution. Let us consider a homogeneous lattice with quantum gates $\hat{C}_{n,t}(\theta_1)$, in which a general noise process deviates some quantum gates from their ideal operation, leading them to effectively behave like $\hat{C}_{n,t}(\theta_2)$. We explored both the spatial and the temporal inhomogeneous scenarios, in which we assume discrete-time quantum walks effectively ruled by two distinct quantum gates, just like in Refs.~\cite{PhysRevLett.93.190503,PhysRevE.96.012111,PhysRevA.89.042307}. We show noise correlations driving quantum walks with spatial inhomogeneities from the well-known exponentially localized (stochastic and uncorrelated noise) to the superdiffusive spreading. This scenario displays an exciting performance in which the superdiffusive exponent is almost unvarying to the inhomogeneity degree $\Delta \theta$. A superdiffusive asymptotic behavior is also reported for quantum walks undergoing temporal noise correlation, contrasting with the diffusive regime exhibited when noise is random and uncorrelated. However, results show the superdiffusive spreading unreachable for specific quantum gate settings.

\section{model}
\label{model}

We consider quantum walks in one-dimensional lattices of interconnected sites. The quantum walker state $\ket{\Psi}$ belongs to a Hilbert space $H=H_{P}\otimes H_{C}$, where $H_{C}$ is a complex vector space of dimension two associated with the internal degree of freedom, here spanned in the basis $\{|\uparrow\rangle,|\downarrow\rangle\}$. The position Hilbert space $H_{P}$ is spanned by the basis $\{|n\rangle\}$ with the lattice nodes $n \in \mathbb{Z}$.

Each step of evolution consists of quantum gates $\hat{C}_{n,t}$ located in the lattice sites, which act on the quantum walker and shuffle its internal state. Such a state establishes the spatial redistribution to be performed by the shift operator $\hat{S}$. Thus, starting from an initial state $\ket{\Psi_{t=0}}$, the dynamical evolution is accomplished by recursively applying the unitary transformation $\ket{\Psi_{t+1}}=\hat{U}\ket{\Psi_t}$, with $\hat{U}=\hat{S}\cdot(\hat{C}_{n,t}\otimes I_{P})$.

An arbitrary quantum walker state in the $t$-th time step is written as 
\begin{eqnarray}
\ket{\Psi_t} = \sum_n \big(\psi_{n,t}^{\uparrow}\ket{\uparrow} 
+ \psi_{n,t}^{\downarrow}\ket{\downarrow}\big)\otimes\ket{n},
\end{eqnarray}
in which $\psi_{n,t}^{\uparrow}$ and $\psi_{n,t}^{\downarrow}$ are complex amplitudes that satisfy $\sum_n \norma{\psi_{n,t}^{\uparrow}} + \norma{\psi_{n,t}^{\downarrow}} =1$. Quantum gates $\hat{C}_{n,t}$, the ones responsible for mixing the internal degree of freedom, are arbitrary SU (2) unitary operators given by 
\begin{eqnarray}\label{quantum_coin_operator}
\hat{C}_{n,t}(\theta)=\cos[\theta_{n,t}]\hat{Z} + \sin[\theta_{n,t}]\hat{X}, 
\end{eqnarray}
with $\theta_{n,t} \in [0,2\pi]$. $\hat{Z}, \hat{X}$ are the Pauli matrices and $I_{P}$ describes the identity operator in space of positions. The subindices $n$ and $t$ indicate the possible spatial and temporal dependencies of these quantum gates, respectively. The stepping of the quantum walker to the left and right is achieved by using the shift operator $\hat{S}$, given as
\begin{eqnarray}\label{shift_operator}
\hspace{-.3cm}\hat{S}=\sum_n \left(\ket{n+1}\bra{n}\otimes\ket{\uparrow}\bra{\uparrow}+\ket{n-1}\bra{n}\otimes\ket{\downarrow}\bra{\downarrow}\right).
\end{eqnarray}


It is well-known that such quantum walk protocol, with single and steady quantum gates, provides the asymptotical behavior of ballistic spreading (except for $\theta=\pi/2$, for which the particle remains confined). This scenario is modified by disturbances on the quantum gates~\cite{VenegasAndraca2012}. Consider a homogeneous lattice with quantum gates $\hat{C}_{n,t}(\theta_1)$, where a general noise process $\mathcal{D}$ deviates some quantum gates from their ideal operation, causing them to effectively behave like $\hat{C}_{n,t}(\theta_2)=\mathcal{D}\hat{C}_{n,t}(\theta_1)$. Such error processes can be unitary, resulting from over- or under-rotation in qubit control pulses~\cite{PhysRevLett.117.170502}. Assuming a local noise, where interferences act on individual quantum gates located along the lattice positions, the lattice displays spatial inhomogeneity, where quantum gates $\hat{C}_{n,t} (\theta_2)$ emerge for some sites $n$ of the lattice. In the absence of a temporal change of quantum gates ($\hat{C}_{n,t} = \hat{C}_{n}$), we effectively deal with a random spatial arrangement of $\hat{C}_{n} (\theta_1)$ and $\hat{C}_{n}(\theta_2)$, where $\hat{C}_{n}(\theta_2)$ appears according to the percentage $p$ and $\hat{C}_{n}(\theta_1)$ with ($1 - p$). Results show that such a scenario exhibits an exponentially localized quantum walk for any $\hat{C}_{n}(\theta_2)$ other than $\hat{C}_{n}(\theta_1)$~\cite{PhysRevLett.106.180403,crespi2013}. However, when all quantum gates simultaneously feel the same disturbances, but in randomly specific time steps along the time evolution, we tackle a temporal inhomogeneity ($\hat{C}_{n,t} = \hat{C}_{t}$). Thus, $\hat{C}_{t}(\theta_2)$ appears throughout the time evolution according to the percentage $p$ and $\hat{C}_{t}(\theta_1)$ with ($1 - p$). Results report a diffusive spreading for any $\hat{C}_{n}(\theta_2)$ other than $\hat{C}_{n}(\theta_1)$~\cite{PhysRevA.89.042307}.

Here, we introduce a minimally biased noise model exhibiting a short-range correlation. We assume the previously described random distribution, adding the condition that $\hat{C}_{n,t}(\theta_2)$ always appears in pairs. The quantum gates are unchanging over time for a spatial-dependent noise scenario. Therefore, the stochastic arrangement of quantum gates throughout the lattice sites obeys the constraint that $\hat{C}_{n}(\theta_2)$ always appears on adjacent sites. For time-dependent noise, where quantum gates are the same at all lattice sites and undergo random changes at each time step, we assume the constraint that $\hat{C}_{t}(\theta_2)$ invariably appears in two consecutive time steps.

We consider the initial state of the quantum walker to be a symmetric one of the form
\begin{eqnarray}
\ket{\Psi_{t=0}}=\frac{1}{\sqrt{2}}(\ket{\uparrow} + i \ket{\downarrow})\otimes |n_0\rangle, 
\end{eqnarray}
with the initial position $n_0$ of the quantum walker defined at the central site of the lattice. We consider open chains as the boundary condition throughout the analysis, with large enough lattice sizes so that the wave function does not reach the edges over the time course described. Considering the stochastic nature and the uniqueness of each sample owing to its specific noise or inhomogeneity, averaging over multiple samples provides a more representative and robust perspective on the system. Thus, we establish an ensemble of fifty subsequent and independent quantum walks to evaluate its average behavior.

\section{Results and discussion}
\label{Results_and_discussion}  
\subsection{Spatial noise}

\begin{figure}[h]
    \centering
            \resizebox{7.1cm}{5.4cm}{\includegraphics{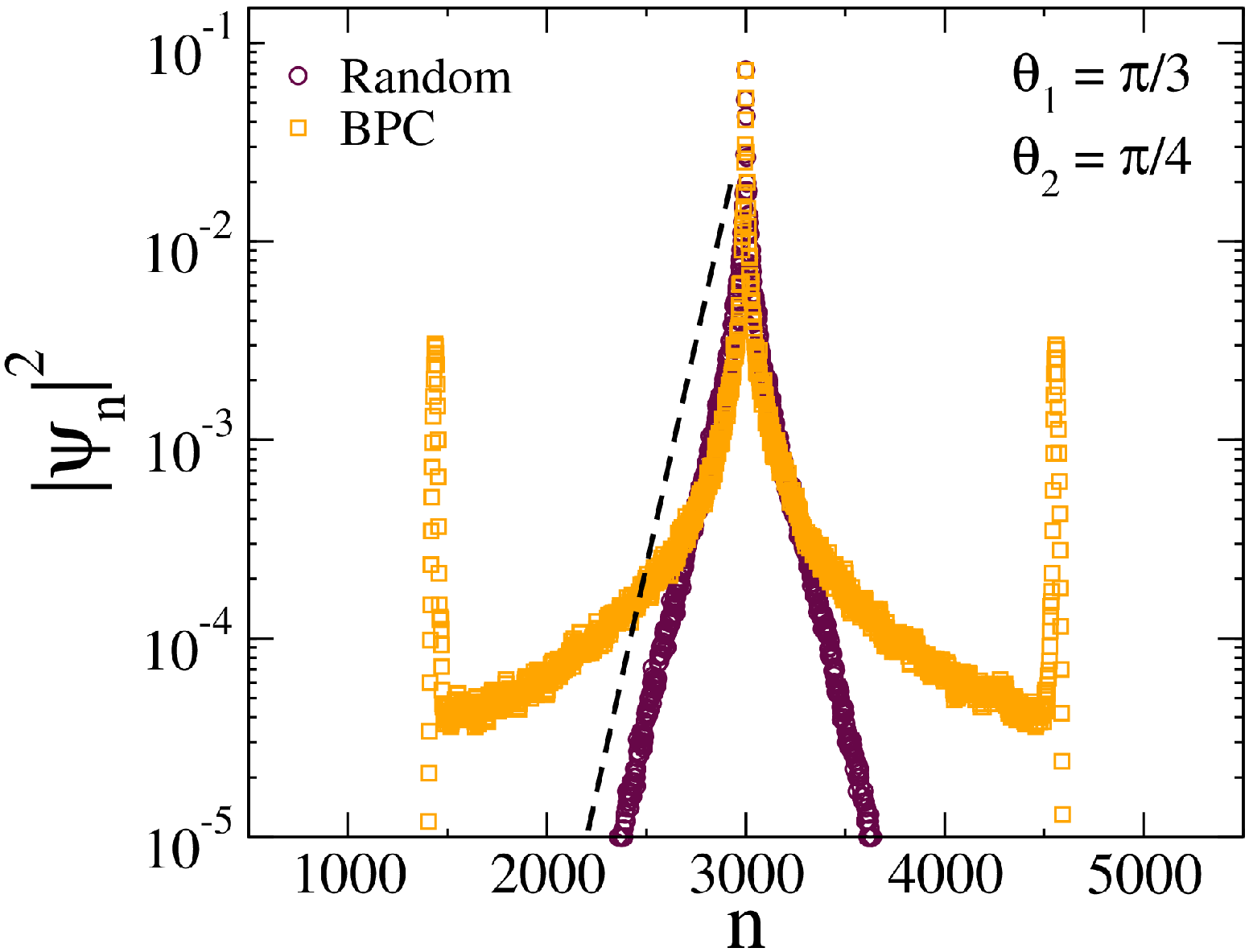}}
    \caption{Average probability distributions after 3000 time-steps for a quantum walker subjected to uncorrelated (brown circles) and spatially correlated noise (orange squares). In the absence of correlations, the quantum walker's profile exhibits a signature of Anderson localization, characterized by exponential decay and linear fitting in the semilog scaled plot (the dashed line is a guide for eyes). With binary pair correlations (BPC), the wave function spreads further, and the exponentially decaying tails give way to wave-packet fronts exhibiting sharp cutoff, which suggests a delocalized behavior.}
    \label{fig1}
\end{figure}

Using the numerical method described above, we start by examining the weight of the proposed correlation over the asymptotic behavior of the quantum walker. In Fig.~\ref{fig1}, we show a snapshot of average probability distribution profiles, taking as reference the scenario of quantum gates arranged randomly and independently. We set $\theta_1 = \pi/3$. In the absence of correlations, we consider $\theta_2 = \pi/4$ stochastically distributed along the lattice sites with a proportion of $p=0.1$. Under the condition of binary pair correlation (BPC), the lattice exhibits the Hadamard quantum gates ($\theta_2$) distributed in pairs (adjacent sites) with a proportion of $p=0.1$. In the absence of correlations, we observe the probability distribution of the quantum walker strictly around the initial position. Wave-function profile exhibiting an exponential decay exposes a signature of Anderson localization, which corroborates the Refs.~\cite{PhysRev.109.1492,PhysRevLett.106.180403,crespi2013}. A distinct scenario is described by systems with binary pair correlations since the wave function is no longer concentrated around the starting position. Now, exponentially decaying tails give way to wave-packet fronts exhibiting sharp cutoff, suggesting a delocalized spreading regime.

\begin{figure}[t]
    \centering
      \resizebox{6.8cm}{8.27cm}{\includegraphics{Fig2.eps}}  
    \caption{Average standard deviation of the quantum walker distribution vs. time for noiseless, random, and binary pair-correlated quantum walks. (a) $\theta_1=\pi/3$ and $\theta_2=\pi/4$ and (b) $\theta_1=4\pi/15$ and $\theta_2=\pi/4$. An asymptotic superdiffusive behavior emerges from the binary pair correlation, contrasting with the characteristic localized regime exhibited by quantum walks subjected to uncorrelated random noise.}
    \label{fig2}
\end{figure}

To better understand the previous results, we follow the time evolution of the wave-function spreading by using the standard deviation 
\begin{equation}
\sigma(t)=\sqrt{\langle n^2(t)\rangle-\langle n(t)\rangle^2},
\label{eq:sd}
\end{equation}
where {\small $\langle n^2(t)\rangle=\sum_n n^2|\Psi_n(t)|^2$} and  {\small $\langle n(t)\rangle=\sum_n n|\Psi_n(t)|^2$}. Its characteristic power-law behavior $\sigma(t)\sim t^\alpha$ quantifies the spreading properties of wave-function, as the ballistic ($\alpha=1.0$), the diffusive ($\alpha=0.5$) and the localized ($\alpha=0.0$) behavior. In Fig.~\ref{fig2}a, we explore $\theta_1 = \pi/3$ and $\theta_2 = \pi/4$, the same $\theta$ settings employed in Fig.~\ref{fig1}. We also consider a noiseless quantum walk as a reference, which displays a ballistic spreading evident from the linear growth of $\sigma(t)$ as time evolves. Such behavior contrasts with the localized quantum walk verified when $\theta_1$ and $\theta_2$  are randomly and independently spatial distributed, characterized by $\sigma(t)$ saturating after an initial transient, i.e., $\sigma(t)\sim t^{0}$. The main point is the superdiffusive spreading for quantum walks subjected to binary pair correlations, characterized by exponent $\alpha \approx {0.74}$. The constancy of the asymptotic spreading performance, even with replacing the quantum gate $\theta_1$ from $\pi/3$ to $4\pi/15$ (see Fig.~\ref{fig2}b) or differing scenarios of inhomogeneity ($p = 0.1, 0.3, 0.5$), highlights the robustness and generalization of the finding.

\begin{figure}[t!]
    \centering
            \resizebox{8.1cm}{11.7cm}{\includegraphics{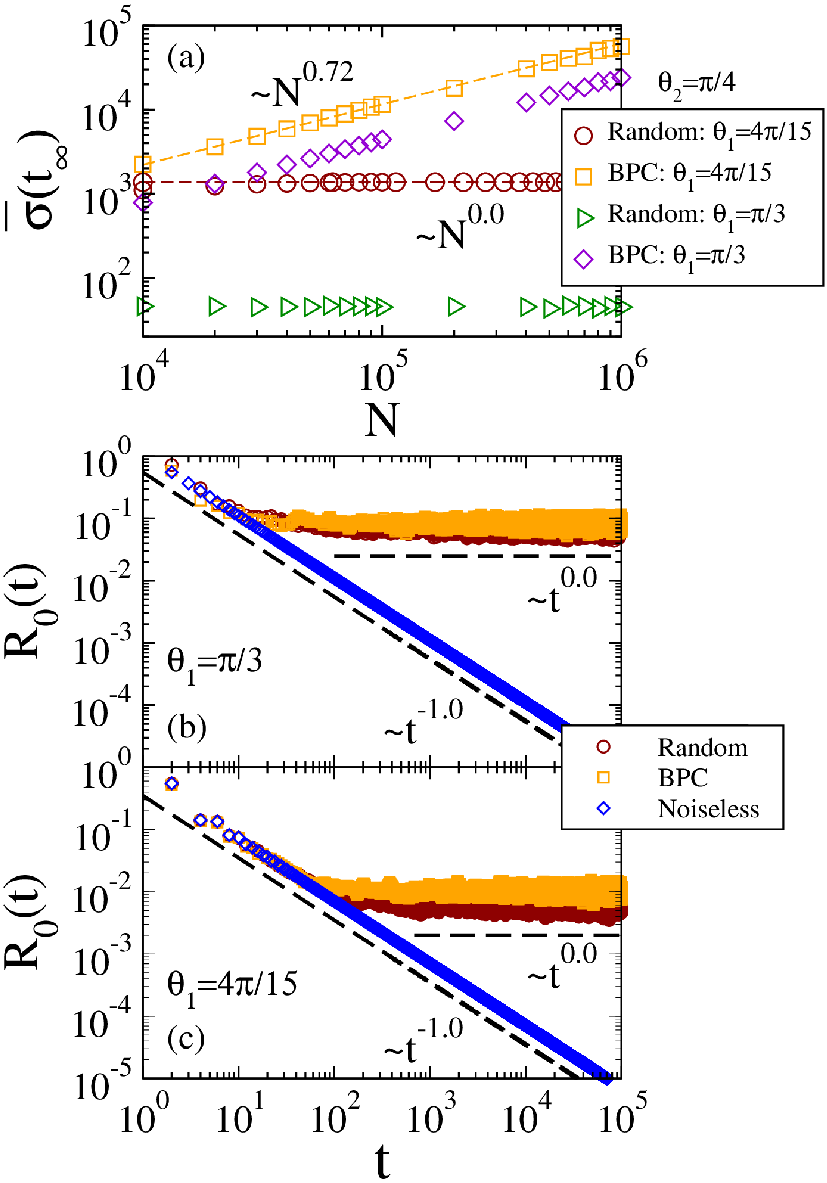}}    
    \caption{(a) The finite-size scaling computed for the long-time average of $\sigma(t)$ supports the previous findings, which unveils the superdiffusive behavior for lattices with binary pair correlation. However, the persistence of a nonvanishing return probability at long times (see b-c) reveals that a fraction of the walker remains localized around its initial position.}
    \label{fig3}
\end{figure}

The supplementary analysis in Fig.~\ref{fig3} examines the lattice size dependence for the wave-packet width in a long-time regime and the time evolution of return probability. Considering the initial transient behavior, which becomes more prolonged as the error $\Delta \theta=|\theta_1-\theta_2|$ decreases, we explore in Fig.~\ref{fig3}a lattice sizes ranging from $N = 10.000$ to $N = 1.000.000$ sites using the previously defined quantum gates: $\theta_2 = \pi/4$ with $\theta_1 = 4\pi/15$ and $\theta_2 = \pi/4$ with $\theta_1 = \pi/3$. By adopting $p=0.5$ as the benchmark value from now on, we observe systems with a stochastic and uncorrelated spatial ordering of quantum gates displaying a size-independent scenario [$\overline{\sigma}(t_\infty) \sim N^0$], indicative of the localized regime. Conversely, finite-size scaling for systems with binary pair-correlated spatial noise reveals $\overline{\sigma}(t_\infty)\sim N^{0.72}$, consistent with the superdiffusive spreading reported earlier. Such results fully agree with previous findings in Fig.~\ref{fig2}. However, the return probability 
\begin{equation}
R_0(t) = \sum_{\alpha = \uparrow,\downarrow} |\langle n_0|\otimes\langle \alpha|\Psi_n(t)\rangle|^2,
\end{equation}
which indicates the probability of the walker returning to the initial position $n_0$ at time $t$, reveals that a fraction of the walker remains localized around its initial position even over an extended time evolution. This behavior contrasts with a fully delocalized regime, such as the noiseless quantum walk, where the return probability tends to zero at long times~\cite{Konno2010,Xu2010,PhysRevA.101.023802}.

\begin{figure}[t!]
    \centering
            \resizebox{6.8cm}{10.2cm}{\includegraphics{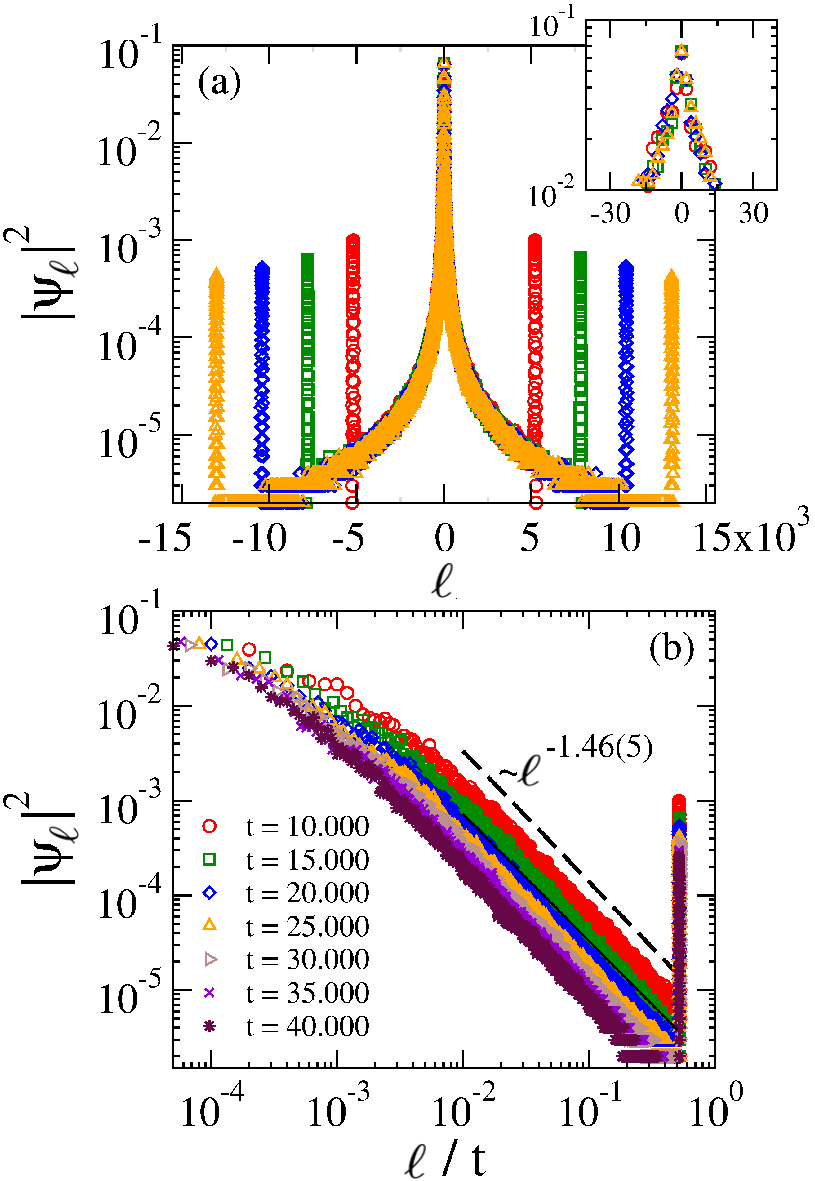}}    
    \caption{Profile of the average probability distributions along the lattice sites ($\ell = n - n_0$) computed at different time steps $t$ for a quantum walker subjected to spatially correlated noise. Quantum gates are the same employed in Fig.~\ref{fig1}. (a) Despite the concentration around the initial position $n_0$ (the magnified view shown in the inset confirms the nonvanishing return probability), the wavefront advances as time evolves. (b) Analysis of the distribution at distinct evolution times shows such wavefront advancing ballistically ($\sim t$) and a power-law tail $|\Psi_n|^2 \sim \ell^{-\varphi}$, with scaling exponent $\varphi = 1.46(5)$.}
    \label{fig4}
\end{figure}

To better understand this intriguing scenario, we evaluate the average probability distributions along the lattice sites ($\ell = n - n_0$) at different time steps $t$ for a quantum walker evolving on lattices with binary pair-correlated spatial noise, as depicted in Fig.~\ref{fig4} . The quantum gates are set to  $\theta_1 = \pi/3$ and $\theta_2 = \pi/4$, identical to those used in Fig.~\ref{fig1}. Despite the concentration around the initial position $n_0$, we observe in Fig.~\ref{fig4}a the wavefront advancing across the lattice sites over time. The inset provides a magnified view around the starting position, with the collapsed data confirming the persistent nonvanishing return probability as time evolves, consistent with previous findings. In Fig.~\ref{fig4}b, analysis of the probability distributions at different evolution times reveals the ballistic advancement ($\sim t$) of the wavefront and a power-law tail $|\Psi_n|^2 \sim \ell^{-\varphi}$, characterized by a scaling exponent $\varphi = 1.46(5)$.  This power-law tail extends up to a cutoff distance $\ell_m$ from the initial position, corresponding to the wavefront. These temporal and spatial scaling behaviors enable the evaluation of the wave-packet mean-square displacement by
\begin{equation}
\sigma^2(t) = \sigma^2(\ell_0) + \sum_{\ell_0}^{\ell_m(t)}\ell^2\left(|\Psi_0|^2\ell^{-\varphi}\right).
\end{equation}
Here, $\ell_0$ denotes the characteristic distance beyond which the power-law decay occurs, and $|\Psi_0|^2$ represents the coefficient of the asymptotic power-law decay of the wave-packet. Considering that $\sigma^2(t) \sim \sum_{\ell_0}^{\ell_m(t)}|\Psi_0|^2\ell^{-(\varphi-2)}$ and $\varphi - 2 < 1$, the wave-packet mean-square displacement is observed to be sensitive to the wave-packet cutoff. Specifically, this series yields in the long-time regime $\sigma^2 \sim \ell_m^{1.54}$. Since the wavefront advances ballistically, we can infer that $\sigma \sim t^{0.77}$, which supports the previously reported superdiffusive behavior.
\begin{figure}[t]
    \centering
            \resizebox{7.cm}{11.7cm}{\includegraphics{Fig5.eps}}    
    \caption{Asymptotic exponent ($\alpha$) of standard deviation and long-time average of return probability for different quantum gates $\theta_1$, with (a-b) $\theta_2=\pi/4$ and (c-d) $\theta_2=\pi/3$. Binary pair correlation induces a notable transition from exponential localization to superdiffusive spreading. Despite correlated noise, a fraction of the walker remains localized around the initial position, a phenomenon not exclusive to Hadamard quantum gates appearing in pairs.}
    \label{fig5}
\end{figure}

The question that arises is whether this superdiffusive behavior persists for other quantum gates or if alternative quantum gates could potentially induce diffusive or subdiffusive spreading. The answer is presented in Fig.~\ref{fig5}, which explores the asymptotic exponent $\alpha$ of characteristic power-law $\sigma(t) \sim t^\alpha$, and the average return probability at long times. These quantities are examined as a function of $\theta_1$ while taking into account systems with quantum gates arranged unbiasedly or with binary pair correlation. We explore (a-b) $\theta_2=\pi/4$ and (c-d) $\theta_2=\pi/3$. Fig.~\ref{fig5}a extends the previous results to further $\theta_1$ quantum gates and confirms a dominant superdiffusive regime for quantum walks with binary pair-correlated noise. The same behavior is observed when other quantum gates play the role of binary pair-correlated (see Fig.~\ref{fig5}c). The asymptotic exponent $\alpha$ stands approximately unvarying ($\sim 0.73$) when noise is correlated, even for a small $\Delta \theta$. On the other hand, uncorrelated spatial noise leads to a stagnation of the spread after an initial transient and hence a $\sigma(t)\sim t^0$ in the asymptotic regime, in entire agreement with Ref.~\cite{PhysRevLett.106.180403,crespi2013,PhysRev.109.1492}. We observe exceptions for quantum gates $\theta_1$ equivalent to the pair-correlated quantum gates $\theta_2$, in which the fully extended (ballistic) regime is achieved. On the other hand, employment of Pauli X quantum gates results in a localized behavior, coming from swapping the amplitudes of states $|\uparrow\rangle$ and $|\downarrow\rangle$ that corresponds to a negation operation. Despite the alteration in the prevailing spreading regime, the binary-pair correlation has no significant effect on the return probability of the quantum walker. After evaluating different configurations of $\theta_1$ and $\theta_2$, results indicate a nonvanishing return probability at long-time evolutions, i.e., $R_0(t) \sim t^{-\beta}$ becomes $R_0 (t_\infty) \sim t^{-0}$. Results depicted in Fig.~\ref{fig5}(b, d) indicate that a fraction of the walker persists in its initial position in the presence or absence of correlation in spatial noise even at long-time evolutions, undermining the conception of complete delocalization.

These results can be understood when we look at the quantum gates playing the role of altering the probability of the quantum walker (qubit) moving to the right or left. With all quantum gates identical, the system is translation invariant, and the generalized eigenfunctions are described by Bloch waves, infinitely extended over the whole lattice. However, the system is no longer translationally invariant when quantum gates vary randomly in space. The sequence of such quantum gates becomes reflective and decoherent for a walker trying to spread through the respective lattice, which inhibits spreading due to interference effects between multiple scatterings of the qubit wave function, causing the eigenfunctions to become exponentially localized~\cite{PhysRev.109.1492,PhysRevLett.106.180403,crespi2013}. In agreement with previous works over that class of inhomogeneities~\cite{PhysRevLett.65.88,doi:10.1126/science.252.5014.1805,PhysRevLett.82.2159}, the noise with binary pair-correlation gives rise to extended states, that emerge as transparent (resonant) states for small finite samples (domains) and contribute to the spreading of the quantum walker on the lattice.  Not all modes are extended and effectively sensitive to the emergence of such domains, resulting in a fraction of the quantum walker stuck around the initial position~\cite{PhysRevResearch.2.023411}. Although a similar phenomenology has been reported in electronic transport~\cite{PhysRevLett.65.88}, we observe particular features such as the absence of superdiffusive, diffusive, or localized regimes, with thresholds between them depending on the inhomogeneity degree. Our results suggest a prevailing superdiffusive scenario, asymptotically independent of $\Delta \theta$. This scenario helps us understand the superdiffusive quantum walks reported for Thue-Morse- and Fibonacci-type spatial inhomogeneities~\cite{PhysRevE.96.012111}, where binary pairs spontaneously appear throughout the sequencing.

\subsection{Temporal noise}

\begin{figure}[!b]
    \centering
      \resizebox{6.8cm}{8.27cm}{\includegraphics{Fig6.eps}}  
    \caption{Average standard deviation of the quantum walker distribution vs. time for noiseless, random, and binary pair-correlated quantum walks. (a) $\theta_1=\pi/3$ and $\theta_2=\pi/4$ and (b) $\theta_1=4\pi/15$ and $\theta_2=\pi/4$. An asymptotic superdiffusive behavior emerges from the binary pair correlation, contrasting with the characteristic diffusive regime exhibited by quantum walks subjected to uncorrelated random noise.}
    \label{fig6}
\end{figure}

This section is devoted to studying walks with time-dependent varying quantum gates. We start by looking at the standard deviation (see Eq.~\ref{eq:sd}), establishing a comparative analysis between systems with binary pair correlations and their uncorrelated counterparts. In Fig.~\ref{fig6} we evaluate the time evolution by considering $\theta_2=\pi/4$ with (a) $\theta_1=\pi/3$ and (b) $\theta_1=4\pi/15$. We observe noiseless quantum walks displaying a ballistic behavior, characterized by power law $\sigma(t)\sim t$. On the other hand, systems in which $\theta_1$ and $\theta_2$ appear randomly and independently exhibit a diffusive character, corroborating previous studies~\cite{PhysRevE.96.012111}. Our analysis unveils the emergence of superdiffusive quantum walks resulting from the binary pair correlations. This superdiffusive character is maintained even when the percentage of appearance of the $\theta_2$ gates ($p$) is changed. Additionally, such regime seems to be independent of how far $\theta_2$ is from $\theta_1$, as the superdiffusive behavior is the same for $\theta_1 - \theta_2 = \pi/12$ and $\theta_1 - \theta_2 = \pi/60$.

\begin{figure}
    \centering
      \resizebox{8.1cm}{11.7cm}{\includegraphics{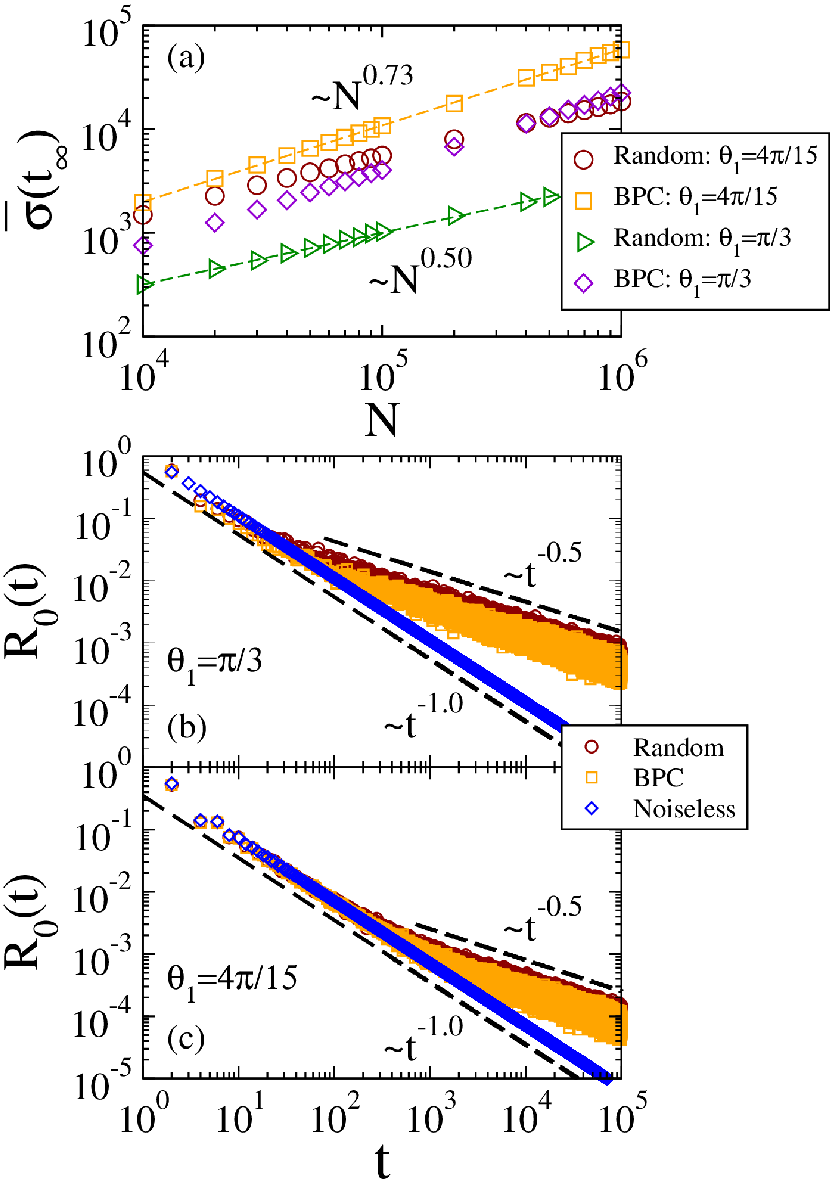}}  
    \caption{(a) Investigating the long-term average of $\sigma(t)$ using finite-size scaling supports the previous findings, which reveal superdiffusive behavior in lattices featuring binary pair correlation. However, the temporal evolution of the return probability demonstrates that the observed performance remains coherent with systems subjected to uncorrelated random noise, following a power law $R_0(t)\sim t^{-0.5}$ (see b-c).}
    \label{fig7}
\end{figure}

The previous behavior is reinforced by analysis of finite-size scaling of the wave-packet width in an asymptotic regime (see Fig.~\ref{fig7}a). By employing quantum gates as depicted in Fig.~\ref{fig6} for lattice sizes ranging from $N=10.000$ to $N=1.000.000$, we observe agreeing results. Systems with quantum gates undergoing an uncorrelated alternation exhibit a diffusive character, identified by $\overline{\sigma}(t_\infty)\sim N^{0.5}$, whereas systems with binary pair correlations exhibit a superdiffusive behavior ($\sim N^{0.73}$). Additionally, we explore in Fig.~\ref{fig7}b-c the probability of the walker returning to its initial position $n_0$ at time $t$. A behavior change is observed in the presence of noise, leaving the $R_0(t)\sim t^{-1.0}$ of noiseless quantum walks~\cite{Konno2010,Xu2010,PhysRevA.101.023802} to $R_0(t)\sim t^{-0.5}$, whether the noise is uncorrelated or correlated. As observed in the spatial noise scenario, correlation in noise plays a relevant role in the dynamics by promoting resonant extended states that drive a ballistic advance on the wavefront and promote a power-law tail on the wave packet. Such change in the distribution tail is fundamental for manifesting the reported superdiffusive scenario. Not all states are effectively sensitive to the binary pair correlation, thus preserving the dynamic profile around the initial position.

\begin{figure}
    \centering
            \resizebox{7.cm}{11.7cm}{\includegraphics{Fig8.eps}}  
    \caption{The asymptotic exponent of the standard deviation ($\alpha$) and the return probability ($\beta$) are examined for various quantum gates $\theta_1$, with (a-b) $\theta_2=\pi/4$ and (c-d) $\theta_2=\pi/3$. Binary pair correlation induces a transition in the dominant spreading behavior from diffusive to superdiffusive. Exceptions around $\theta_1=2\pi/3$ suggest that certain quantum gates are insensitive to binary pair correlation. Altering the spreading regime does not notably impact the time evolution of the return probability.}
    \label{fig8}
\end{figure}

In order to know whether such superdiffusive behavior also extends when other quantum gates are involved, we examine the power-law exponents $\alpha$ [$\sigma(t) \sim t^\alpha$] and $\beta$ [$R_0(t) \sim t^{-\beta}$] as a function of $\theta_1 $, taking into account the quantum gates $\theta_2 = \pi/4$ and $\theta_2 = \pi/3$ (see Fig.~\ref{fig8}). As a reference, we also present data for the regimen without correlations, named random. We observed a predominant superdiffusive spreading, whether the dimerized quantum gates are Hadamard (a) or $\theta_2=\pi/3$ (c). This scenario is present even for the Pauli-X quantum gates, whose bit-flip character between $|\uparrow\rangle$ and $|\downarrow\rangle$ has proved to be dominant when the noise interfered only in the spatial arrangement of the quantum gates. However, we observed an anomalous behavior when $\theta_1$ is in the vicinity of $2\pi/3$ while evaluating on both $\theta_2$. Our observations suggest that these quantum gates are insensitive to binary pair correlation. Notably, the significant change in the spreading regime does not reflect in the asymptotic return probability (as illustrated in Fig.~\ref{fig8}b,d). Data indicates that there is no significant difference between correlated or uncorrelated temporal noise. Both systems exhibit an approximately constant exponent $\beta$ (around $\beta \approx 0.5$), regardless of how much $\theta_2$ differs from $\theta_1$.



\begin{figure}
    \centering
            \resizebox{8.6cm}{7.1cm}{\includegraphics{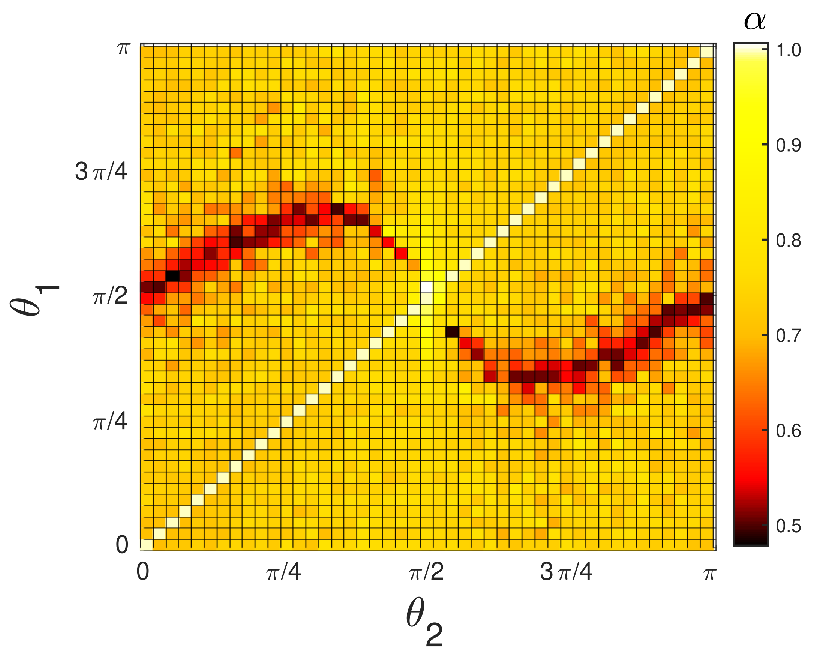}}  
    \caption{Density plot illustrating the characteristic asymptotic exponent of the standard deviation ($\alpha$) in the plane of $\theta_2$ and $\theta_1$ quantum gates. Binary pair correlation supports a superdiffusive spreading for most evaluated quantum gates. Notably, specific combinations of quantum gates remain unresponsive to noise correlation, preserving the diffusive regime despite its existence.}
    \label{fig9}
\end{figure}

To better characterize this unusual regime, we show in Fig.~\ref{fig9} a density plot of asymptotic exponent characteristic of the standard deviation ($\alpha$) in a plane of $\theta_2$ vs $\theta_1$ quantum gates. The scenario where $\theta_1$ and $\theta_2$ are identical corresponds to a noiseless system, resulting in ballistic wave function spreading. Such behavior is exhibited by $\alpha=1.0$ along the diagonal. To improve the clarity of the density plot, we excluded the exceptional case where both $\theta_1$ and $\theta_2$ are configured as Pauli-X gates, which results in a well-known localized behavior~\cite{Nielsen_2000,PhysRevE.100.032106}. The most significant observation is the prevalence of superdiffusive behavior across a wide range of quantum gates $\theta_1$ and $\theta_2$, highlighting the influence of correlated noise on wave-function spreading. The sequential repetition of a single quantum gate over time promotes the spreading of the quantum walker. However, introducing different quantum gates disrupts the phase relationship between different wave function components, interfering with the superposition and interference of states, thus compromising the wave function spreading. The emergence of temporal binary pair correlation mitigates this loss of phase coherence in some components, thereby contributing to the observed superdiffusive behavior. However, the temporal alternation between quantum gates $\theta_1$ and $\theta_2$ and the subjacent effects of interference and superposition of the quantum walker states prove effectively unvarying to the presence of binary pair correlations for particular combinations of quantum gates, sustaining diffusive spreading. The occurrence of this singular scenario depends on the specific values of $\theta_1$, described by an approximate relationship $\theta_1 \approx \pi/2[1-\sin(2\theta_2-\phi)/4]$ where $\phi=\pi$. This scenario points to the possibility of particular quantum gates performing as a filter of correlations, which can contribute to the progress of tools and algorithms for quantum processes under noise influence~\cite{PhysRevLett.127.170403}.

\section{Summary and concluding remarks}\label{sec:conclusions}

In summary, we have studied the transport properties in discrete-time quantum walks undergoing a noise correlation. By considering a relaxation in the uncorrelated stochastic noise premiss, we thought the emergence of binary pair-correlated in the random distribution. We have explored spatial and temporal noise scenarios, always drawing a comparative analysis with systems under uncorrelated noise. The dynamics of the quantum walker were computed from a sample mean of independent noises. In systems with spatial inhomogeneity, we observe the binary pair correlation driving the quantum walks from the exponentially localized regime (coming from the stochastic and uncorrelated noise) to superdiffusive spreading. Such behavior holds maintained regardless of the difference between quantum gates $\theta_1$ and $\theta_2$, either by analyzing the time-asymptotic regime, as well as the finite-size scaling, which has unveiled a superdiffusive exponent almost unvarying to the degree of inhomogeneity. Despite this superdiffusive spreading, we identified that a fraction of the walker remains localized around its initial position even after a long-time evolution. Such interesting behavior is allied to the emergence of resonant states induced by correlation in noise. These states facilitate a ballistic advance of the wavefront and contribute to developing a power-law tail in the wave packet distribution. Analytical results reveal that this change in the distribution profile underpins the numerically reported superdiffusive scenario, which is consistent with the findings of Ref.~\cite{de_Moura_2011}.

The binary pair correlation also favors the spreading in the temporal scenario. In such systems, the superdiffusive spreading also emerges from the binary pair correlation, taking the place of the diffusive quantum walks observed for an independent and random temporal inhomogeneity. However, some quantum gates exhibit a remarkable effect of insensitivity to correlations, which seems attractive for studying correlation filters for quantum processes~\cite{PhysRevLett.127.170403}. Our results bring new aspects about superdiffusive quantum walks and the relationship with possible correlations in their protocol, as reported in quantum walks with Fibonacci- or Thue Morse-type aperiodic inhomogeneities~\cite{PhysRevE.96.012111}, where binary pairs appear spontaneously throughout the sequence.  

Our findings provide additional understanding into the emergence of resonant extended modes in noisy systems and how wave packet dispersion unveils their existence. Analytical approaches and entanglement studies represent promising avenues for elucidating further insights into the central theme. To conclude, recent experimental achievement in a time-multiplexing system based on an unbalanced Mach-Zehnder interferometer with a feedback loop~\cite{PhysRevResearch.3.023052} makes us believe that the proposed scheme here is feasible for prompt implementation. Such setup has been proven capable of controlling quantum gates over space and time, designing inhomogeneities.

\section{Acknowledgments}

We are grateful for the insightful discussion with M. L. Lyra. This work was partially supported by CAPES (Coordena\c{c}\~ao de Aperfei\c{c}oamento de Pessoal do N\'ivel Superior), CNPq (Conselho Nacional de Densenvolvimento Cient\'ifico e Tecnol\'ogico), and FAPEAL (Funda\c{c}\~ao de Apoio \`a Pesquisa do Estado de Alagoas).

\bibliographystyle{nature}
\bibliography{refer}

\end{document}